\documentclass[final,3p,times,twocolumn]{elsarticle}

\usepackage{amssymb}

\usepackage{geometry} 
\usepackage{graphicx}  
\usepackage{txfonts}   
\usepackage[colorlinks,linkcolor=blue,anchorcolor=blue,citecolor=blue,urlcolor=blue]{hyperref}
\usepackage{bm}
\usepackage[mathlines]{lineno}
\usepackage{color,soul}
\newcommand{\modR}[1] {{#1}}
\usepackage{natbib}   
\biboptions{sort&compress}

\journal{Carbon}
\begin{document}


\begin{frontmatter}



\title{Catalytic Formation of Narrow Nb Nanowires inside Carbon Nanotubes}



\author{Dan~Liu} %

\author{David Tom\'{a}nek%
\corref{cor}} %
\ead{tomanek@pa.msu.edu}

\address{Physics and Astronomy Department,
            Michigan State University,
            East Lansing, Michigan 48824, USA}
\cortext[cor]{Corresponding author} %


\begin{abstract}
We propose a previously unexplored way to form Nb nanowires from
NbCl$_3$ molecules inside carbon nanotubes (CNTs). We have studied
this reaction by {\em ab initio} density functional calculations
and found it to be catalytically promoted in presence of graphitic
carbon. Our results suggest that chemisorption of NbCl$_3$ on the
CNT is accompanied by a charge transfer of ${\approx}0.5$
electrons to the nanotube wall, which significantly weakens the
Nb-Cl bond. We found that the bcc structure of Nb is not affected
by the small diameter of the nanowire inside a CNT. We have also
identified strong covalent bonds between the nanowires and the
surrounding nanotube that are accompanied by a similar charge
transfer of ${\lesssim}0.5$~e from surface Nb atoms to the
nanotube. The large electronic density of states of bulk Nb at
$E_F$ is not changed much in the confined geometry, suggesting
that ultra-narrow nanowires of Nb may keep their superconducting
behavior and form Josephson junctions in the quasi-1D geometry
while being protected from the ambient by the enclosing CNT
structure.
\end{abstract}




\end{frontmatter}


\section{Introduction}

Niobium holds a special place in the periodic table for several
reasons. It is a ductile transition metal with a high melting
temperature~\cite{Kittel} $T_M=2750$~K and the highest cohesive
energy~\cite{Kittel} $E_{coh}=7.57$~eV/atom in the $4d$ transition
metal series. Nb also displays the highest critical temperature
$T_c=9.25$~K among elemental
superconductors~\cite{{Supercond1963},{Finnemore1966}} and is
unique with its ability to maintain an open body-centered cubic
(bcc) structure until extremely high pressures~\cite{Weck2019} of
hundreds of GPa. Nb is significantly harder than iron and an
important constituent of thermally stable superalloys and
superconducting alloys. There is strong interest in Nb nanowires
covering applications ranging from efficient superconducting
single-photon detectors~\cite{{Prober2010},{Natarajan2012}},
Josephson tunnel junctions~\cite{Henry2014}, and conductive fiber
composites used as supercapacitors~\cite{Mirvakili2015} and energy
storage media~\cite{Mirvakili2013}. A wide range of fabrication
techniques including etching of sputter-deposited
films~\cite{Prober2010} or compounds~\cite{Mirvakili2013},
templated electrodeposition~\cite{Blagg2019}, or deposition on
nano-templates~\cite{Bezryadin03} were only capable of producing
polycrystalline wires with diameters from 10-1000~nm. %
\modR{%
Availability of monocrystalline Nb nanowires with only few
nanometers in diameter is highly desirable especially for quantum
applications in the superconducting regime. Yet the common
synthesis approach based on high-temperature capillarity
filling~\cite{%
{Ajayan93},{Yahachi93},{Tsang94},{Sloan99},{Fan00},{Jeong03},%
{GarciaFuente11},{DT222},{Komsa17}} %
of carbon nanotubes can not be used for Nb due to its high melting
temperature, which is close to the stability limit of carbon
nanotubes. %
} %

Here we propose a previously unexplored approach to synthesize Nb
nanowires with a uniform diameter by %
\modR{%
heating up the NbCl$_3$ molecular crystal  %
}%
in presence of carbon nanotubes (CNTs) with a typical diameter
$d{\lesssim}2$~nm. The nanowires should form by
\modR{%
decomposition of NbCl$_3$ inside CNTs at a temperature well below
the melting point of Nb. In this process, the role of the CNT
is threefold. %
}%
First, the CNT is a template that defines the diameter of the
enclosed. Second, graphitic carbon in the CNT acts as a catalyst,
which lowers the temperature of converting NbCl$_3$ to metallic Nb
and NbCl$_5$ to well below $1,000^\circ$C~\cite{Ripan72}. Finally,
the CNT protects the enclosed Nb nanowire from the ambient. We
have performed {\em ab initio} density functional theory (DFT)
calculations of the reaction
mechanism and found that Nb nanowires may form in an %
\modR{%
activated exothermic process that involves a large number of atoms
in concerted motion. %
} %
The catalytic effect of a CNT originates in its ability to accept
electrons from chemisorbed NbCl$_3$. Our calculations allow us to
identify the optimum geometry of Nb nanowires that form inside the
CNTs. %
\modR{%
Nb nanowires are faceted fragments of the bulk bcc structure with
little reconstruction. They maintain the same high density of
states at $E_F$ as bulk Nb and thus should remain superconducting,
similar to their bulk counterpart. %
}%
More interesting, ultra-narrow Nb nanowire segments, separated by
a narrow gap and held in place by the surrounding CNT, should form
a Josephson junction in the quasi-1D geometry while being
protected from the ambient by the enclosing CNT structure.

At first glance, formation of Nb nanowires inside carbon nanotubes
should not appear surprising, since nanowires of other elements
have been found inside CNTs\cite{%
{Ajayan93},{Yahachi93},{Tsang94},{Sloan99},{Fan00},{Jeong03},%
{GarciaFuente11},{DT222},{Komsa17}}. So far, filling nanotubes by
a metal~\cite{Ajayan93} such as Pb had required initial melting of
this metal, which subsequently entered the nanotubes by capillary
action.
\modR{%
Similar to other refractory metals, this approach fails for Nb
with it melting point~\cite{Kittel} $T_M=2750$~K, which is close
to the stability limit of carbon nanotubes between $951$~K in
presence of oxygen~\cite{Yu2005} and near $3,400$~K in
vacuum~\cite{Wei2011}.
} %

Our proposed approach is very different, as it involves
sublimation of NbCl$_3$ molecules from a molecular crystal, their
entry into the open end of a CNT and subsequent chemisorption on
the inner wall. As we will discuss below, we find that
chemisorption of NbCl$_3$ on the CNT wall is accompanied by a
large charge transfer of $0.5$ electrons from the molecule to the
CNT, which strongly reduces the ionic charge of Cl atoms in this
molecule. The charge redistribution in the adsorbed complex lowers
the energy required to break the Nb-Cl bond by ${\gtrsim}1$~eV.
Consequently, the exothermic conversion of NbCl$_3$ to metallic Nb
and NbCl$_5$ is promoted in presence of graphitic carbon, which
plays the unusual role of a catalyst. There is preliminary
experimental evidence~\cite{YNakanishi-private} that formation of
Nb nanowires inside CNTs from NbCl$_3$ is indeed possible at
temperatures significantly lower than $1,000^\circ$C,
\modR{%
which is required for this reaction to occur in vacuum, without a
catalyst~\cite{Ripan72}. }%

\begin{figure}[t]
\begin{center}
\includegraphics[width=0.90\columnwidth]{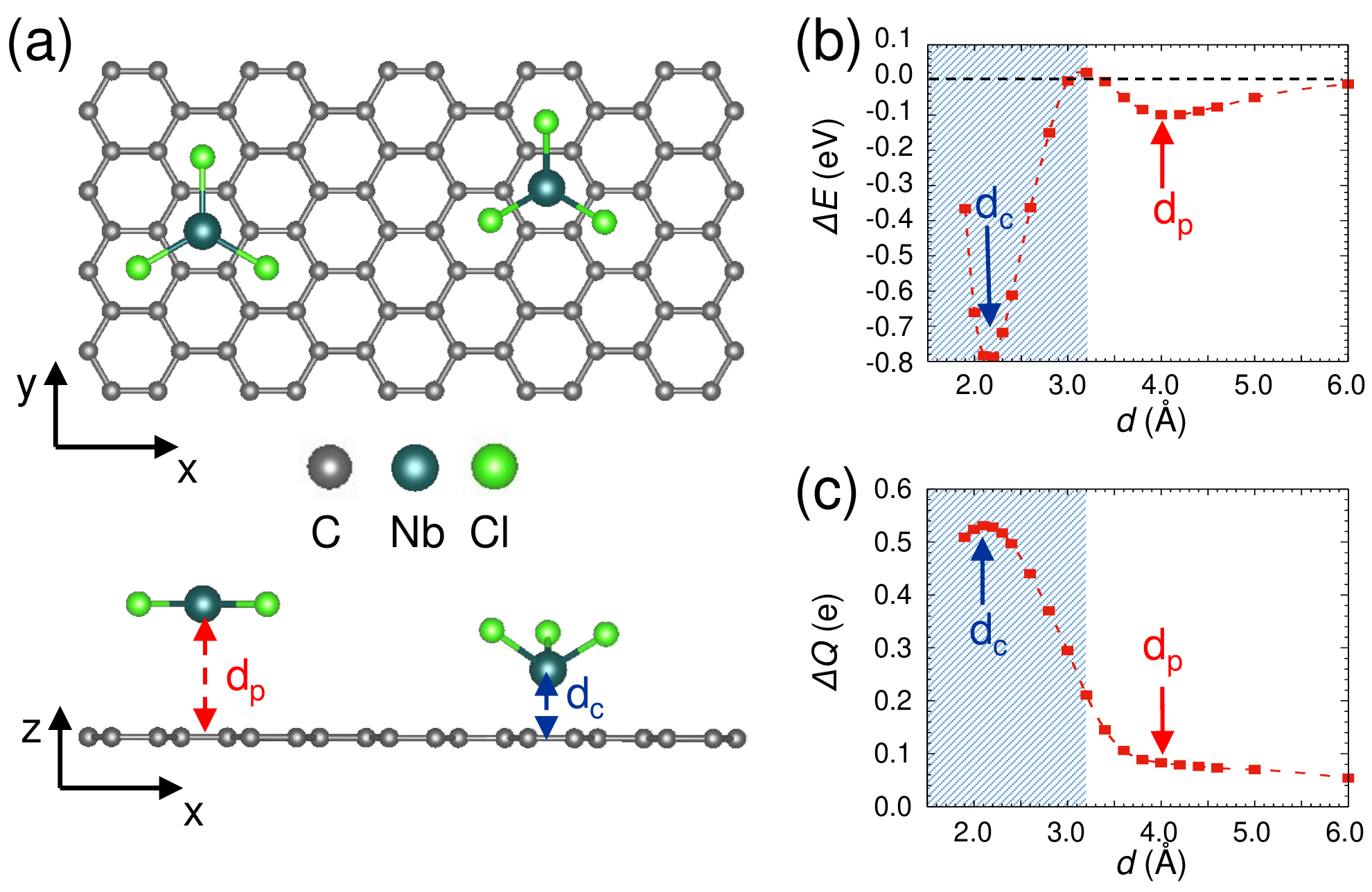}
\end{center}
\caption{Interaction of an NbCl$_3$ molecule with a graphene
         monolayer. %
(a) Side and top views of the molecule at the distance $d=d_c$ and
    $d=d_p$ from graphene. %
(b) Interaction energy ${\Delta}E(d)$ between an NbCl$_3$
    molecule and graphene. %
(c) Charge redistribution upon chemisorption of NbCl$_3$ on
    graphene. %
\label{fig1}}
\end{figure}

\section{Computational results}

We have considered different chlorides of niobium as potential
precursors for metallic Nb and calculated the corresponding
reaction energies by optimizing the initial and the final
structures using {\em ab initio} DFT calculations. Among these, we
found NbCl$_5$ to be unusually stable, as it forms spontaneously
from metallic Nb and chlorine gas according to~\cite{Lavut82}
\begin{equation}
{\rm Nb(s)} + 2.5 {\rm{Cl}}_2{\rm{(g)}} \longrightarrow
{\rm{NbCl}}_5{\rm{(s)}} \,. %
\label{Eq1}
\end{equation}
The observed reaction energy~\cite{{Gross1960},{Lavut82}} of
$-797$~kJ/mol or
$-8.26$~eV per formula unit is in fair agreement with the
calculated value of $-7.48$~eV. The strongly exothermic nature of
this process indicates that the reverse process of forming
metallic Nb from NbCl$_5$ is highly unlikely.

However, we found metallic Nb to form as one of the products of
the decomposition of NbCl$_3$ according
to~\cite{{Ripan72},{Harjanto05}}
\begin{equation}
5{\rm{NbCl}}_{3}{\rm{(s)}} \longrightarrow %
3{\rm{NbCl}}_{5} + 2{\rm{Nb(s)}} \;.%
\label{Eq2}
\end{equation}
\modR{%
This reaction is exothermic and has been reported to occur at
$1,000^\circ$C under vacuum
conditions~\cite{{Ripan72},{Harjanto05}}. Even though NbCl$_{5}$
molecules are listed as one of the products, these molecules are
known to dimerize spontaneously to Nb$_{2}$Cl$_{10}$. Then,
Eq.~(\ref{Eq2}) changes to
\begin{equation}
10{\rm{NbCl}}_3{\rm{(s)}} \longrightarrow %
3{\rm{Nb}}_{2}{\rm{Cl}}_{10}{\rm{(g)}} + 4{\rm{Nb(s)}} \;.%
\label{Eq3}
\end{equation}
The calculated reaction energy of this exothermic process is
${\Delta}E=-19.82$~eV. Even though we know that this reaction does
occur at $1,000^\circ$C, there is only limited information
available about it due to its complexity~\cite{Korzynski18}, which
involves continuous changes in the oxidation state of Nb. The
reaction involves a concerted motion of 40 atoms in a
120-dimensional configuration space that is hard to search.
Consequently, the reaction coordinate and transition states can
not be characterized well, as we expand on in the Supplementary
Material (SM). %
}%

\modR{ %
According to Eq.~(\ref{Eq3}), we consider NbCl$_3$ as a suitable
precursor for the formation of metallic Nb. We note that this
favorable precursor may also form %
} %
by exposing NbCl$_5$ to hydrogen atmosphere according
to~\cite{Harjanto05}
\begin{equation}
{\rm{NbCl}}_5{\rm{(g)}} + {\rm{H}}_2{\rm{(g)}}  \longrightarrow
{\rm{NbCl}}_3{\rm{(s)}} + 2 {\rm{HCl(g)}} \;.%
\label{Eq4}
\end{equation}
This reaction is only slightly exothermic, with the reported
reaction energy~\cite{Harjanto05} of $-35$~kJ/mol.
\modR{ %
Even though we are unable to characterize the reaction coordinate,
transition states and activation energies for the reaction in
Eq.~(\ref{Eq3}), we are still able to shed light on key steps in
the reaction, which involves breaking of some and formation of
other bonds in the Nb-Cl system, leading ultimately to metallic
Nb inside a CNT. %
} %

In the first step, an NbCl$_3$ molecule adsorbs on the inner wall
of the CNT, which we represent by a graphene monolayer in view of
the low curvature of the $2$-nm wide CNT. Our results are
summarized in Fig.~\ref{fig1}. Corresponding results for the
interaction of NbCl$_5$ with graphene are presented in the SM.

As seen in Fig.~\ref{fig1}(a), NbCl$_3$ is a planar molecule that
prefers adsorption in the hollow site on graphene. The adsorption
energetics is depicted in Fig.~\ref{fig1}(b). NbCl$_3$ initially
physisorbs at the distance $d_p=4.07$~{\AA}, gaining $0.10$~eV in
physisorption energy. Only a small energy barrier separates it
from a much more stable chemisorbed state associated with an
energy gain of $0.80$~eV and a much smaller distance
$d_c=2.15$~{\AA} from graphene. As shown in Fig.~\ref{fig1}(a),
the molecule deforms during the chemisorption, bringing the Nb
atom closer to the surface. To better understand the changes in
chemical bonding associated with the adsorption of NbCl$_3$ on
graphene, we performed a single$-\zeta$ Mulliken population
analysis of the system and monitored the net charge $Q$ of
NbCl$_3$ at different values of the adsorption distance $d$. We
then defined the electron depletion ${\Delta}Q(d)$ on the molecule
at distance $d$ with respect to a desorbed molecule at
$d=10$~{\AA} by ${\Delta}Q(d)=Q(d=10$~{\AA}$)-Q(d)$. Our results
in Fig.~\ref{fig1}(c) indicate that the molecule remains almost
neutral down to $d=d_p$, but loses up to $0.53$ electrons as it
approaches graphene in the chemisorbed state. From this viewpoint,
graphene is not inert with respect to the chemical behavior of
NbCl$_3$.

\modR{%
As mentioned above, formation of metallic Nb from NbCl$_3$
molecules during the complex reaction behind Eq.~(\ref{Eq3})
necessitates breaking or rearranging Nb-Cl bonds. Whereas
separating a Cl atom from a free NbCl$_3$ molecule requires
$4.57$~eV, the same process requires only $2.60$~eV when NbCl$_3$
is chemisorbed on graphene or inside a CNT. Also a related
endothermic reaction that is likely to occur during the complex
conversion process, where NbCl$_3$ first dimerizes to Nb$_2$Cl$_6$
and subsequently dissociates into NbCl$_2$ and NbCl$_4$, requires
$1.26$~eV in vacuum and a significantly lower energy of $0.52$~eV
while adsorbed on graphitic carbon of a nanotube. Details of these
particular reactions are
presented in the SM. %
}%

\modR{ %
We attribute the significant decrease of the reaction energy
barriers on graphitic carbon to the electron transfer from the
chemisorbed molecule to the carbon nanotube. This catalytic effect
is unusual, since with few exceptions~\cite{{Gu17},{Xu17}},
graphitic carbon has been known to be catalytically inactive
unless combined with another substance to form a graphene-based
catalyst or a supported graphene catalyst. Consequently, formation
of metallic Nb from NbCl$_3$ according to the reaction in
Eq.~(\ref{Eq3}) should occur inside a CNT at temperatures
significantly lower than the reported value of $1,000^\circ$C in
vacuum~\cite{{Ripan72},{Harjanto05}}. %
} %
This conjecture appears to be supported by preliminary
experimental results~\cite{YNakanishi-private}.


Having established the possibility of forming metallic Nb inside a
CNT, we now have to consider likely structures of finite-diameter
Nb nanowires %
\modR{%
that are dynamically stable. %
}%
Unlike in macroscopic structures including large-diameter wires,
atomic packing may be significantly changed in nanostructures
including nanowires due to the significant effect of surface
tension that exerts compressive stress on the entire structure. We
summarize observed and calculated values of the surface tension
$\gamma$ in Table~\ref{table1}. We have estimated the equivalent
pressure for Nb nanowires in excess of $2$~nm diameter and found
it insufficient to convert the open bcc structure of bulk Nb to a
more close-packed structure, in agreement with previous
results~\cite{Weck2019}. We also cut out and optimized selected
nanowires from bulk Nb(bcc) and found no signs of significant
reconstruction.
\modR{%
Thus, we believe that same as in the bulk, atoms in Nb nanowires
are packed in the bcc structure, with possible atomic relaxations
at the surface. The nanowires discussed in the following are to be
considered stable dynamically, not only statically, since atomic
rearrangement to a more stable structure has been excluded. Still,
there is need to discuss possible shape changes at their faceted
surface inside the nanotube void, which we do in the following. %
}%

%
%

Whereas the most stable structure would minimize the total surface
energy in vacuum, it should minimize the total interface energy
between Nb and the CNT in the Nb@CNT system. To obtain insight
into the different components of the interface energy, we do not
constrain our study to particular nanotube diameters, but rather
study different Nb surfaces and their interaction with a graphene
monolayer.

\begin{figure}[t]
\begin{center}
\includegraphics[width=0.90\columnwidth]{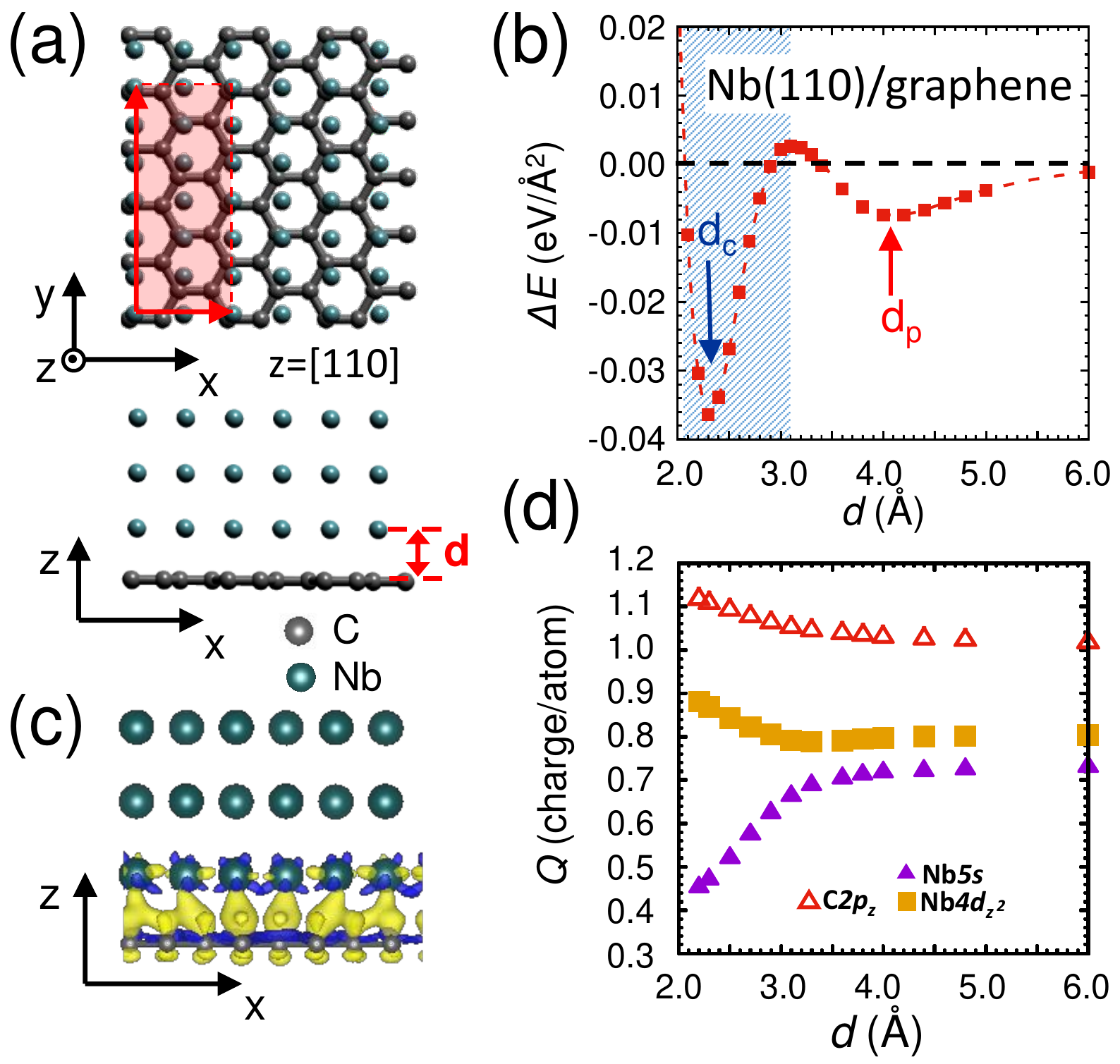}
\end{center}
\caption{Interaction of the Nb(110) surface with a graphene
         monolayer. %
(a) Top and side views of the atomic arrangement, with the
    supercell highlighted by red. %
(b) Interaction energy ${\Delta}E$ between the Nb(110) surface and
    graphene as a function of the separation distance $d$. %
(c) Charge redistribution caused by the interaction of Nb(110)
    with graphene, superposed with the atomic structure. %
    ${\Delta}\rho=\rho$(Nb/graphene)$-\rho$(Nb)$-\rho$(graphene)
    is shown by isosurfaces bounding regions of electron excess at
    $+4{\times}10^{-3}$~e/{\AA}$^3$ (yellow) and electron
    deficiency at $-4{\times}10^{-3}$~e/{\AA}$^3$ (blue). %
(d) Mulliken charge of specific atomic orbitals as a function of
    $d$. %
\label{fig2}}
\end{figure}

\begin{table}[b!]
\caption{%
Characterization of selected Nb surfaces and their interaction
with graphene based on DFT-PBE calculations. $\gamma$ is the
surface energy. The optimum Nb-graphene distance is $d_{c}$ for
chemisorption and $d_{p}$ for physisorption. ${\Delta}E(d_{c})$ is
the Nb-graphene interaction energy and ${\Delta}Q(d_{c})$ is the
electron charge transferred from Nb to graphene at $d_{c}$, both
given per area unit.
}%
\setlength\tabcolsep{3 pt}
\begin{tabular}{lccccc} %
\hline \hline
   \textrm{\ } %
 & \textrm{$\gamma$}
 & \textrm{$d_{c}$} %
 & \textrm{$d_{p}$} %
 & \textrm{$\Delta E(d_{c})$} %
 & \textrm{$\Delta Q(d_{c})$} %
 \\
   \textrm{Surface} %
 & \textrm{(eV/{\AA}$^2$)} %
 & \textrm{({\AA})} %
 & \textrm{({\AA})} %
 & \textrm{(eV/{\AA}$^2$)} %
 & \textrm{(e/{\AA}$^2$)} %
 \\
\hline%
  {(100)} %
& {0.152} %
& {2.32} %
& {-} %
& {0.035} %
& {0.040} %
\\
  {\ } %
& {$0.142^a$} %
& {} %
& {} %
& {} %
& {} %
\\
  {(110)} %
& {0.127} %
& {2.31} %
& {4.07} %
& {0.037} %
& {0.048} %
\\
  {\ } %
& {$0.129^a$} %
& {} %
& {} %
& {} %
& {} %
\\
  {\ } %
& {$0.155^b$} %
& {} %
& {} %
& {} %
& {} %
\\
  {\ } %
& {$0.159^c$} %
& {} %
& {} %
& {} %
& {} %
\\
  {(111)} %
& {0.150} %
& {2.28} %
& {-} %
& {0.022} %
& {0.031} %
\\
  {\ } %
& {$0.146^a$} %
& {} %
& {} %
& {} %
& {} %
\\
  {(121)} %
& {0.146} %
& {2.37} %
& {-} %
& {0.025} %
& {0.032} %
\\
  {\ } %
& {$0.146^a$} %
& {} %
& {} %
& {} %
& {} %
\\
\hline \hline %
$^a$ Ref. \cite{Tran16}.\\
$^b$ Ref. \cite{Mills06}.\\
$^c$ Ref. \cite{Hodkin70}.\\
\end{tabular}
\label{table1}
\end{table}

As a likely candidate for segments of the nanowire surface, we
first consider the Nb(110) surface with the lowest value of
$\gamma$ according to Table~\ref{table1}. We would expect Nb
nanowires with a surface consisting only of (110) facets to be
most stable in vacuum. The interaction of the Nb(110) surface with
graphene, representing a finite-diameter nanowire enclosed in a
CNT, is presented in Fig.~\ref{fig2}. Corresponding results for
the Nb(100), Nb(111) and Nb(121) surfaces are presented in the SM.

We represent the Nb(110) surface by a 3-layer slab. Since Nb(110)
is not epitaxial with the honeycomb lattice of graphene, we
considered a large Nb$_{18}$C$_{16}$ supercell depicted in
Fig.~\ref{fig2}(a), which contained 16 C and 6 Nb atoms at the
interface. Epitaxy, needed for the calculation, was enforced by
compressing Nb(110) by $0.8$\% along the $x-$direction and
stretching it by $2.1$\% along the $y-$direction, whereas graphene
was left in its optimum structure. The Nb(110)/graphene
interaction energy ${\Delta}E$ is plotted in Fig.~\ref{fig2}(b) as
a function of the interface distance $d$. Similar to the NbCl$_3$
molecule on graphene discussed above, we find two adsorption
minima for this system. As also listed in Table~\ref{table1}, the
shallower minimum, which we associate with physisorption, is
characterized by $d_p=4.07$~{\AA} and
${\Delta}E_p=-0.008$~eV/{\AA}$^2$. The deeper minimum, which we
associate with chemisorption, is characterized by $d_c=2.31$~{\AA}
and ${\Delta}E_c=-0.037$~eV/{\AA}$^2$. These interaction energies
are smaller than the surface energies listed in
Table~\ref{table1}, but can not be ignored when discussing the
\modR{ %
surface structure of nanowires inside a CNT. %
}%

We find similarities in the interaction of graphene with the
surface of Nb and a NbCl$_3$ molecule. Unlike in the physisorbed
state of Nb on graphene, there is significant electron transfer
from Nb to graphene in the more stable chemisorbed state. The
spatial extent of this charge transfer is, however, limited to the
one Nb layer closest to graphene, as shown in Fig.~\ref{fig2}(c).
To better understand the details of this charge transfer, we
plotted the Mulliken population charge, obtained using a
single-$\zeta$ basis, in Fig.~\ref{fig2}(d) for atomic orbitals
that were most affected. In graphene, it was the population of the
C$2p_z$ orbital that increased at smaller interlayer distances. In
Nb, the population of the Nb$4d_{z^2}$ state increased and that of
the Nb$5s$ state decreased significantly at small interlayer
distances. This confirms that the most significant bonding between
Nb and C is associated with $pd\sigma$ bonding gaining strength
over the $sp\sigma$ bond as Nb and C atoms approach each other.
This bonding character is also reflected in the charge density
redistribution in Fig.~\ref{fig2}(c).

\begin{figure}[t]
\begin{center}
\includegraphics[width=1.0\columnwidth]{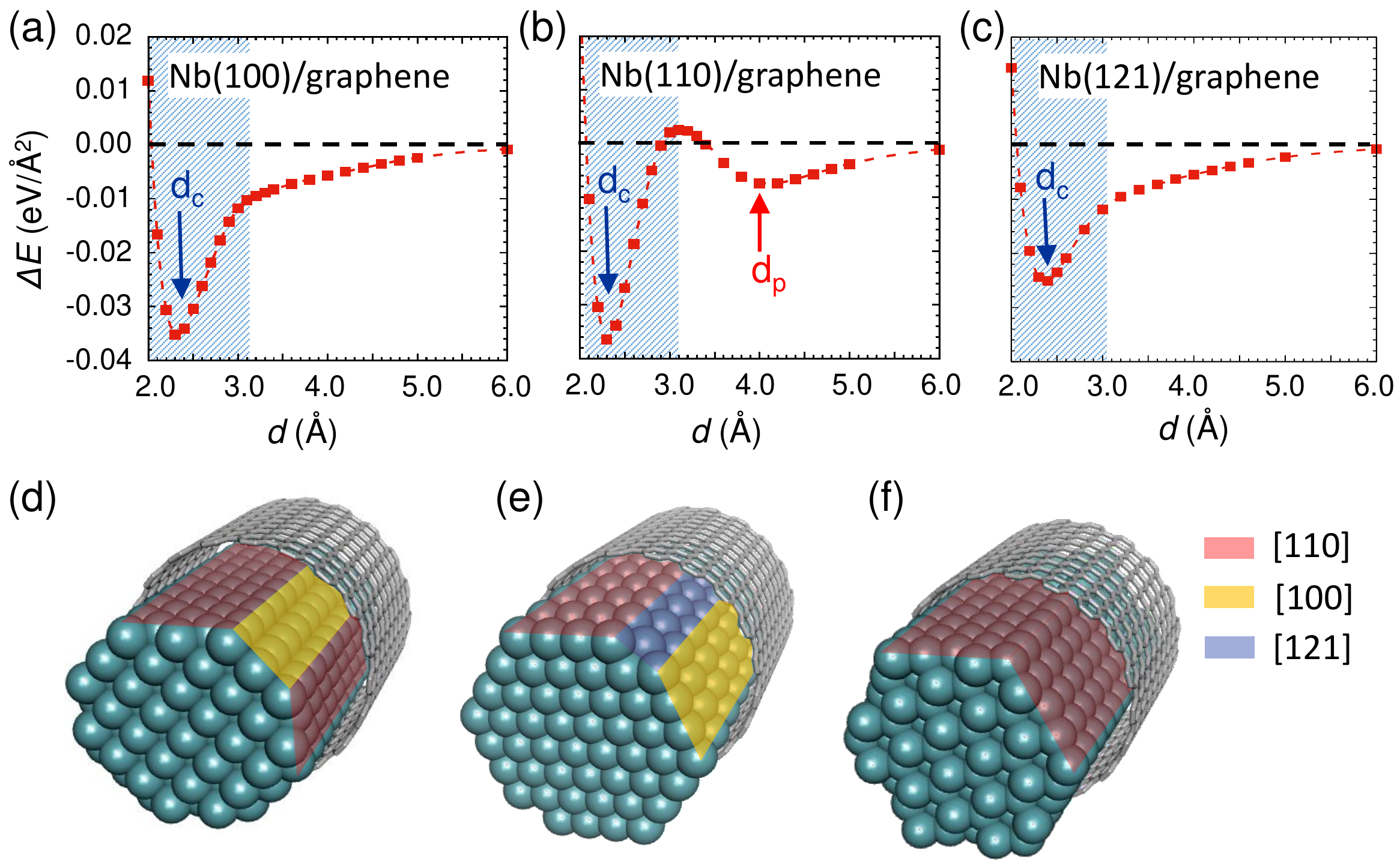}
\end{center}
\caption{Interaction energy of %
(a) Nb(100), %
(b) Nb(110), and %
(c) Nb(121) %
with a graphene monolayer. These surfaces form facets on
monocrystalline Nb nanowires. Most stable Nb nanowire
geometries are shown schematically for nanowires with %
(d) (100) and (110), %
(e) (110), (121) and (100), and %
(f) only (110) facets. %
\label{fig3}}
\end{figure}

The interaction energies of different Nb surfaces with graphene
are compared in Figs.~\ref{fig3}(a)-\ref{fig3}(c). Unlike for the
(110) surface discussed above, there are no physisorption minima
for Nb(100) and Nb(121) interacting with graphene. Among the three
surfaces considered, Nb(110) interacts most strongly with
graphene. The numerical values for the interaction energies of the
different surfaces with graphene are summarized in
Table~\ref{table1}.

Candidate structures for the most stable, ${\approx}2$~nm-wide Nb
nanowires are shown in Figs.~\ref{fig3}(d)-\ref{fig3}(f).
Of these, the structure in Fig.~\ref{fig3}(f) with a hexagonal
cross-section is the most stable in vacuum, as it is terminated by
[110] facets only. Since the Nb(110) surface interacts most
strongly with graphene, this structure is favored even more
strongly inside a CNT on energy grounds. In nanowires shown in
Figs.~\ref{fig3}(d) and \ref{fig3}(e), less stable [100] and [121]
facets coexist with [110] facets, making these structures
energetically less favorable in vacuum than that in
Fig.~\ref{fig3}(f). Since also the interaction of [100] and [121]
facets with graphene is weaker than that of [110] facets, these
structures remain energetically less favorable with respect to the
structure in Fig.~\ref{fig3}(f) also inside a CNT. We thus focus
on the preferred structure in Fig.~\ref{fig3}(f) in the following.

\begin{figure}[t]
\begin{center}
\includegraphics[width=1.0\columnwidth]{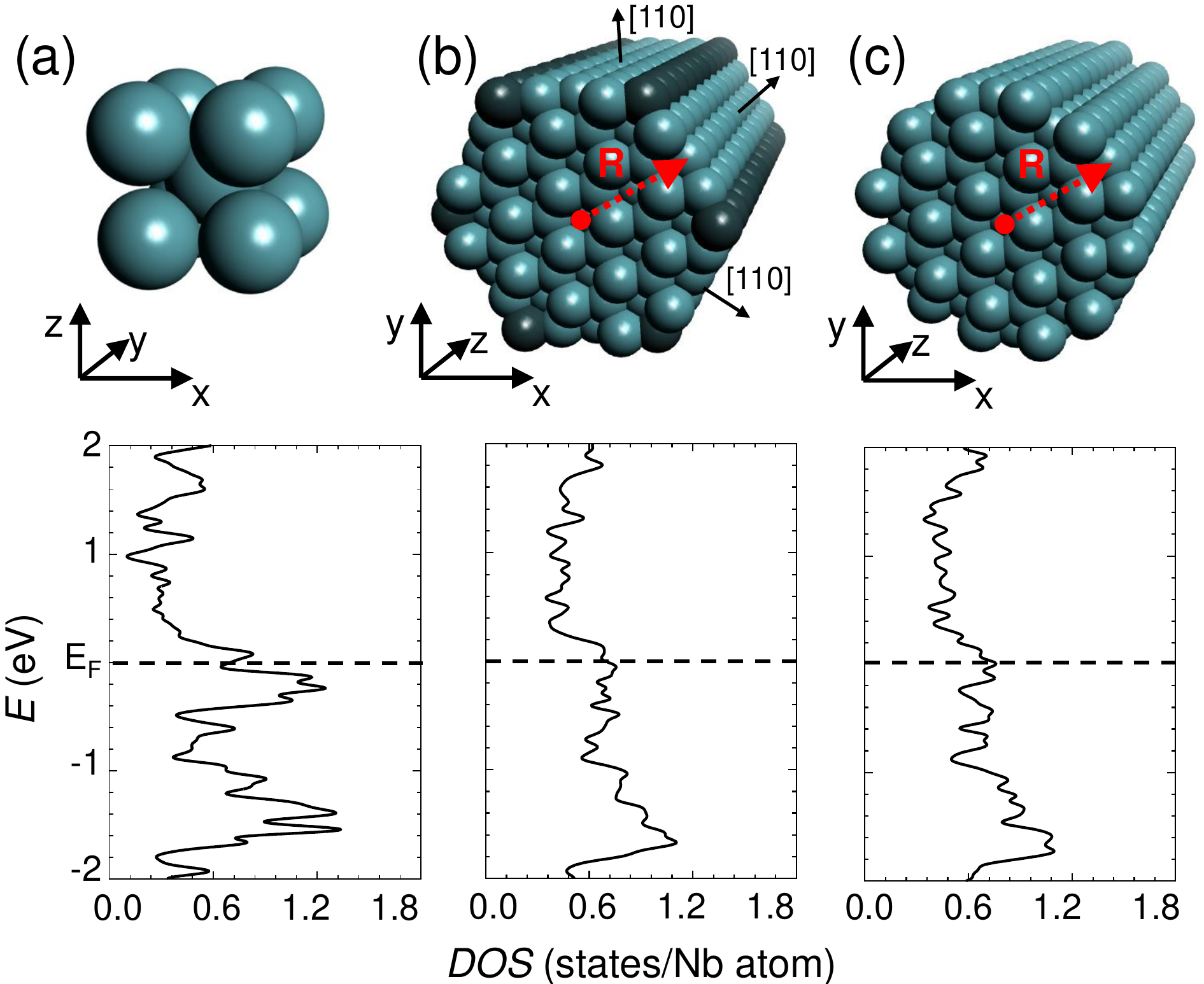}
\end{center}
\caption{Electronic density of states (DOS) of %
(a) bulk Nb(bcc), %
(b) a Nb nanowire terminated by (110) facets, and %
(c) the Nb nanowire of (b) with atoms at the edges %
    of the facets eliminated. %
\label{fig4}}
\end{figure}

Finally, to estimate transport properties of Nb nanowires, we
compare their electronic density of states (DOS) $N(E)$ to the
infinite bcc structure in Fig.~\ref{fig4}. We display
schematically the bulk bcc structure in the top panel of
Fig.~\ref{fig4}(a), that of the nanowire of Fig.~\ref{fig3}(f) in
Fig.~\ref{fig4}(b), and the same structure with edge atoms removed
in Fig.~\ref{fig4}(c). The DOS in the bottom panel of
Fig.~\ref{fig4}(a) reveals the metallic character of bulk Nb and
reflects well the partial filling of the Nb$4d$ band. The DOS of
the two nanowires, shown in the bottom panels of
Figs.~\ref{fig4}(b) and \ref{fig4}(c) appears flatter near the
Fermi level. Interestingly, the DOS value at $E_F$, $N(E_F)$, is
essentially the same, possibly slightly higher in the nanowires
than in the bulk structure. This would indicate that the
conductivity of the nanowires should remain comparable to the bulk
structure. We expect an electron transfer from the nanowires to
the enclosing nanotube that involves only the outermost atoms in
the nanowire. Depending on the nanowire diameter, the position of
the Fermi level should be lowered by ${\lesssim}0.1$~eV with no
effect on $N(E_F)$. The CNTs surrounding the nanowires may be
intrinsically semiconducting or metallic and will contribute
negligibly to the conductance of the system. In random samples,
$2/3$ of all single-wall CNTs are semiconducting.

Possibly more interesting than conducting behavior is the expected
change in the critical temperature $T_c$ for superconductivity. In
BCS superconductors including Nb, $T_c$ can be estimated using the
McMillan equation~\cite{McMillan68}. %
%
%
The key quantities determining $T_c$ are the the Debye frequency
$\omega_D$ and the electron-phonon coupling constant $\lambda$.
Since the structure of the nanowires is essentially the same as
that of their bulk counterpart, the vibrational spectra and thus
$\omega_D$ should be the same. More important are changes in
$\lambda$, which appears in the exponent of the expression for
$T_c$. One factor in $\lambda$, namely the Bardeen-Pines
interaction describing electron scattering by a phonon, should be
the same in Nb nanowires and bulk Nb due to their similar
electronic and phonon structure. The other factor in $\lambda$ is
$N(E_F)$, which we found to be essentially the same, if not
higher, in the nanowires of Nb as in the bulk. Thus, we expect Nb
nanowires down to $2$~nm diameter to remain superconducting, with
the same or possibly higher value of $T_c$ than the $9.25$~K
value~\cite{{Supercond1963},{Finnemore1966}} reported for bulk Nb.

\section{Discussion}

The CNTs play a triple role in our study. Besides their catalytic
activity to lower the activation barrier for the decomposition of
NbCl$_3$ and their function as a template to form metallic Nb
nanowires, they also protect these nanowires from the ambient.

\modR{%
A proper characterization of a chemical reaction involves
identifying the reaction coordinate, the geometry and energy of
transition states, and a detailed discussion of the dynamical
stability of the final state structure. As we mentioned above, the
reaction involves a concerted motion of 40 atoms in a
120-dimensional configuration space, which makes it impossible to
search for the optimum transformation path between the initial and
the final states, which we have characterized. We have also spent
significant effort finding the optimum structure of nanowires that
fill the void inside a given carbon nanotube. Of course, in
absence of this steric constraint, the nanowires would decompose
to the bulk Nb(bcc) metal, the most stable structure in the phase
diagram of Nb. Clearly, many unknowns remain regrading the
reaction details. Still, we have established that due to the
catalytic activity of graphitic carbon, the reaction should run
below $1,000^{\circ}$C, at which it had been observed under vacuum
conditions. %
}%

The complex decomposition reaction of NbCl$_3$ leads not only to
metallic Nb, but also to solid NbCl$_5$. The two bulk allotropes
of NbCl$_5$ are wide-gap semiconductors and do not affect the
electronic properties of the Nb nanowires, as we expand in more
detail in the SM. We have also found that the interaction between
NbCl$_5$ and the surrounding CNT wall is very weak and does not
involve any charge transfer.

In a realistic sample, we may expect that the Nb nanowire may not
be contiguous. Rather, finite-length Nb nanowire segments, which
are held in place by the surrounding nanotube, may be separated by
narrow gaps or short NbCl$_5$ segments along the axial direction
of a CNT. Assuming that the surrounding CNT is semiconducting, the
Nb-vacuum-Nb or Nb-NbCl$_5$-Nb structure may behave as a
nanometer-wide, precisely controllable Josephson junction, the key
component of future quantum computers. Besides its nanometer
diameter, a significant advantage of a Nb-based Josephoson
junction would be its functionality up to $T_c$(Nb)$=9$~K, a
significant advantage over Al-based Josephson junctions that only
operate below $T_c$(Al)$=2$~K.


\section{Summary and Conclusions}

We have proposed a previously unexplored way to form Nb nanowires
inside carbon nanotubes (CNTs)
\modR{ %
using not capillary filling by the metal, but rather filling the
CNTs with NbCl$_3$ molecules, which decompose to metallic Nb and
Nb$_{2}$Cl$_{10}$ molecules. %
We used {\em ab initio} density functional calculations to get
insight into this reaction, which turns out to be quite complex,
as it involves a concerted motion of 40 atoms and Nb changing its
oxidation state continuously. Whereas the reaction coordinate and
associated activation barriers remain obscure, our calculations
confirm that the reaction is strongly exothermic. Our finding that
particular reaction steps are catalytically promoted in presence
of graphitic carbon implies that the reaction temperature should
decrease below $1,000^{\circ}$C required for it to run in vacuum.
The same process could be used to form nanowires of other
refractory metals including W with an even higher melting point.
}%
According to our calculations, chemisorption of NbCl$_3$ on the
CNT is accompanied by a charge transfer of ${\approx}0.5$
electrons to the nanotube wall, which significantly weakens the
Nb-Cl bond. We found that the bcc structure of Nb is not affected
by the small diameter of the nanowire inside a CNT. We have also
identified strong covalent bonds between the nanowires and the
surrounding nanotube that are accompanied by a similar charge
transfer of ${\lesssim}0.5$~e from surface Nb atoms to the
nanotube. We found that the large electronic density of states at
$E_F$ is not changed much in the confined geometry, suggesting
that ultra-narrow nanowires of Nb may keep their superconducting
behavior in the quasi-1D geometry while being protected from the
ambient by the enclosing CNT structure. Under favorable
conditions, Nb nanowire segments, separated by narrow gaps, may be
held in place by the surrounding nanotube and form Josephson
junctions operating at $T{\lesssim}T_c$(Nb)$=9$~K.

\section{Computational Techniques}

Our calculations of the stability, equilibrium structure and
electronic structure have been performed using density functional
theory (DFT) as implemented in the {\textsc{SIESTA}}~\cite{SIESTA}
code. Periodic boundary conditions have been used throughout the
study, with 2D slabs or graphene and Nb represented by a periodic
array of slabs separated by a 20~{\AA} thick vacuum region. We
also used a periodic array to represent 1D Nb nanowires separated
by a 15~{\AA} thick vacuum region. We used the
Perdew-Burke-Ernzerhof (PBE)~\cite{PBE} exchange-correlation
functional. The {\textsc{SIESTA}} calculations used
norm-conserving Troullier-Martins
pseudopotentials~\cite{Troullier91}, a double-$\zeta$ basis
including polarization orbitals, and a mesh cutoff energy of
$250$~Ry to determine the self-consistent charge density, which
provided us with a precision in total energy of
${\lesssim}2$~meV/atom. The reciprocal space has been sampled by a
fine grid~\cite{Monkhorst-Pack76} of $7{\times}3$~$k$-points in
the 2D Brillouin zones (BZ) of the supercells representing
Nb/graphene slabs. The 1D Brillouin zone of Nb nanowires has been
sampled by $15$~$k$-points. Geometries have been optimized using
the conjugate gradient (CG) method~\cite{CGmethod}, until none of
the residual Hellmann-Feynman forces exceeded $10^{-2}$~eV/{\AA}.


\section*{Acknowledgments}

We thank Andrii Kyrylchuk for useful insight into the stability of
Nb salts. We thank Yusuke Nakanishi and Zheng Liu for providing
the initial idea for this study. We further appreciate scientific
discussions about the synthesis of Nb nanowires with Katsumi
Kaneko, Shuwen Wang, and Yung-Chang Lin. D.L. and D.T. acknowledge
financial support by the NSF/AFOSR EFRI 2-DARE grant number
EFMA-1433459. Computational resources have been provided by the
Michigan State University High Performance Computing Center.


\begin{thebibliography}{10}
\expandafter\ifx\csname url\endcsname\relax
  \def\url#1{\texttt{#1}}\fi
\expandafter\ifx\csname
urlprefix\endcsname\relax\def\urlprefix{URL }\fi
\expandafter\ifx\csname href\endcsname\relax
  \def\href#1#2{#2} \def\path#1{#1}\fi

\bibitem{Kittel}
C.~Kittel, Introduction to Solid State Physics, 7th Edition,
Wiley, 2004.

\bibitem{Supercond1963}
B.~T. Matthias, T.~H. Geballe, V.~B. Compton, Superconductivity,
Rev. Mod.
  Phys. 35 (1963) 1--22.

\bibitem{Finnemore1966}
D.~K. Finnemore, T.~F. Stromberg, C.~A. Swenson, Superconducting
properties of
  high-purity niobium, Phys. Rev. 149 (1966) 231--243.

\bibitem{Weck2019}
P.~F. Weck, J.~P. Townsend, K.~R. Cochrane, S.~D. Crockett, N.~W.
Moore, Shock
  compression of niobium from first-principles, J. Appl. Phys. 125~(24) (2019)
  245905.

\bibitem{Prober2010}
A.~J. Annunziata, O.~Quaranta, D.~F. Santavicca, A.~Casaburi,
L.~Frunzio,
  M.~Ejrnaes, M.~J. Rooks, R.~Cristiano, S.~Pagano, A.~Frydman, D.~E. Prober,
  Reset dynamics and latching in niobium superconducting nanowire single-photon
  detectors, J. Appl. Phys. 108~(8) (2010) 084507.

\bibitem{Natarajan2012}
C.~M. Natarajan, M.~G. Tanner, R.~H. Hadfield, Superconducting
nanowire
  single-photon detectors: physics and applications, Supercond. Sci. Technol.
  25~(6) (2012) 063001.

\bibitem{Henry2014}
M.~D. Henry, S.~Wolfley, T.~Monson, R.~Lewis, Ga lithography in
sputtered
  niobium for superconductive micro and nanowires, Appl. Phys. Lett. 105~(7)
  (2014) 072601.

\bibitem{Mirvakili2015}
S.~M. Mirvakili, M.~N. Mirvakili, P.~Englezos, J.~D.~W. Madden,
I.~W. Hunter,
  High-performance supercapacitors from niobium nanowire yarns, ACS Appl.
  Mater. \& Interf. 7~(25) (2015) 13882--13888.

\bibitem{Mirvakili2013}
S.~M. Mirvakili, A.~Pazukha, W.~Sikkema, C.~W. Sinclair, G.~M.
Spinks, R.~H.
  Baughman, J.~D.~W. Madden, Niobium nanowire yarns and their application as
  artificial muscles, Adv. Funct. Mater. 23~(35) (2013) 4311--4316.

\bibitem{Blagg2019}
K.~Blagg, T.~Greymountain, W.~Kern, M.~Singh, Template-based
electrodeposition
  and characterization of niobium nanowires, Electrochem. Commun. 101 (2019)
  39--42.

\bibitem{Bezryadin03}
A.~Rogachev, A.~Bezryadin, Superconducting properties of
polycrystalline nb
  nanowires templated by carbon nanotubes, Appl. Phys. Lett. 83~(3) (2003)
  512--514.

\bibitem{Ajayan93}
P.~M. Ajayan, S.~Iijima, Capillarity-induced filling of carbon
nanotubes,
  Nature 361~(6410) (1993) 333--334.

\bibitem{Yahachi93}
Y.~Saito, T.~Yoshikawa, Bamboo-shaped carbon tube filled partially
with nickel,
  J. Cryst. Growth 134~(1) (1993) 154--156.

\bibitem{Tsang94}
S.~C. Tsang, Y.~K. Chen, P.~J.~F. Harris, M.~L.~H. Green, A simple
chemical
  method of opening and filling carbon nanotubes, Nature 372~(6502) (1994)
  159--162.

\bibitem{Sloan99}
J.~Sloan, D.~M.~Wright, S.~Bailey, G.~Brown, A.~P.~E.~York,
K.~S.~Coleman,
  M.~L.~H.~Green, J.~Sloan, D.~M.~Wright, J.~L.~Hutchison, H.-G. Woo,
  Capillarity and silver nanowire formation observed in single walled carbon
  nanotubes, Chem. Commun. (1999) 699--700.

\bibitem{Fan00}
X.~Fan, E.~C. Dickey, P.~C. Eklund, K.~A. Williams, L.~Grigorian,
R.~Buczko,
  S.~T. Pantelides, S.~J. Pennycook, Atomic arrangement of iodine atoms inside
  single-walled carbon nanotubes, Phys. Rev. Lett. 84 (2000) 4621--4624.

\bibitem{Jeong03}
G.-H. Jeong, R.~Hatakeyama, T.~Hirata, K.~Tohji, K.~Motomiya,
T.~Yaguchi,
  Y.~Kawazoe, Formation and structural observation of cesium encapsulated
  single-walled carbon nanotubes, Chem. Commun. (2003) 152--153.

\bibitem{GarciaFuente11}
A.~Garc{\'{\i}}a-Fuente, V.~M. Garc{\'{\i}}a-Su{\'{a}}rez,
J.~Ferrer, A.~Vega,
  Structure and electronic properties of molybdenum monatomic wires
  encapsulated in carbon nanotubes, J. Phys.: Cond. Matt. 23~(26) (2011)
  265302.

\bibitem{DT222}
T.~Fujimori, R.~dos Santos, T.~Hayashi, M.~Endo, K.~Kaneko,
D.~Tom\'{a}nek,
  Formation and properties of selenium double-helices inside double-wall carbon
  nanotubes: experiment and theory, ACS Nano 7 (2013) 5607--5613.

\bibitem{Komsa17}
H.-P. Komsa, R.~Senga, K.~Suenaga, A.~V. Krasheninnikov,
Structural distortions
  and charge density waves in iodine chains encapsulated inside carbon
  nanotubes, Nano Lett. 17~(6) (2017) 3694--3700.

\bibitem{Ripan72}
R.~Ripan, I.~Ceteanu, Chimia metaleor, Vol.~2, Editura
didactic$\breve{\rm{a}}$
  si pedagogic$\breve{\rm{a}}$, Bucharest, 1966, translated under the title
  {\em Neorganicheskaya khimia}, Moscow, Mir, 1972, vol. 2, p. 189.

\bibitem{Yu2005}
H.~Yu, C.~Lu, T.~Xi, L.~Luo, J.~Ning, C.~Xiang, Thermal
decomposition of the
  carbon nanotube/sio$_2$ precursor powders, J. Therm. Anal. Calorim. 82~(1)
  (2005) 97--101.

\bibitem{Wei2011}
X.~Wei, M.-S. Wang, Y.~Bando, D.~Golberg, Thermal stability of
carbon nanotubes
  probed by anchored tungsten nanoparticles, Sci. Technol. Adv. Mater. 12~(4)
  (2011) 044605.

\bibitem{YNakanishi-private}
Yusuke Nakanishi, Yung-Chang Lin, Zheng Liu, Shuwen Wang, and
Katsumi Kaneko
  (private communication).

\bibitem{Lavut82}
E.~Lavut, B.~Timofeyev, V.~Yuldasheva, G.~Galchenko, Enthalpy of
formation of
  niobium pentachloride, J. Chem. Thermodyn. 14~(6) (1982) 531--535.

\bibitem{Gross1960}
P.~Gross, C.~Hayman, D.~L. Levi, G.~L. Wilson, The heats of
formation of metal
  halides. niobium and tantalum pentachlorides, Trans. Faraday Soc. 56 (1960)
  318--321.

\bibitem{Harjanto05}
S.~Harjanto, A.~Shibayama, K.~Sato, G.~Suzuki, T.~Otomo,
Y.~Takasaki,
  T.~Fujita, Thermal decomposition of nbcl$_5$ in reductive atmosphere by using
  hydrogen gas, Resour. Process. 52~(3) (2005) 113--121.

\bibitem{Korzynski18}
M.~D. Korzy\'{n}ski, L.~Braglia, E.~Borfecchia, C.~Lamberti,
M.~Dinc\v{a},
  Molecular niobium precursors in various oxidation states: An {XAS} case
  study, Inorg. Chem. 57~(21) (2018) 13998--14004.

\bibitem{Gu17}
W.~Gu, J.~Bai, B.~Dong, X.~Zhuang, J.~Zhao, C.~Zhang, J.~Wang,
K.~Shih,
  Catalytic effect of graphene in bioleaching copper from waste printed circuit
  boards by acidithiobacillus ferrooxidans, Hydrometallurgy 171 (2017)
  172--178.

\bibitem{Xu17}
X.~Xu, Y.~Liu, Z.~Liu, F.~Ke, C.~Lin, K.~Liu, Z.~Zhang, Z.~Hu,
X.~Li, X.~Guo,
  Pristine graphene as a catalyst in reactions with organics containing {C=O}
  bonds, arXiv:1707.05607 (2017).

\bibitem{Tran16}
R.~Tran, Z.~Xu, B.~Radhakrishnan, D.~Winston, W.~Sun, K.~A.
Persson, S.~P. Ong,
  Surface energies of elemental crystals, Sci. Data 3 (2016) 160080.

\bibitem{Mills06}
K.~C. Mills, Y.~C. Su, Review of surface tension data for metallic
elements and
  alloys: Part 1-pure metals, Int. Mater. Rev. 51~(6) (2006) 329--351.

\bibitem{Hodkin70}
E.~N. Hodkin, M.~G. Nicholas, D.~M. Poole, The surface energies of
solid
  molybdenum, niobium, tantalum and tungsten, J. Less Common Met. 20~(2) (1970)
  93--103.

\bibitem{McMillan68}
W.~L. McMillan, Transition temperature of strong-coupled
superconductors, Phys.
  Rev. 167 (1968) 331--344.

\bibitem{SIESTA}
E.~Artacho, E.~Anglada, O.~Dieguez, J.~D. Gale, A.~Garcia,
J.~Junquera, R.~M.
  Martin, P.~Ordejon, J.~M. Pruneda, D.~Sanchez-Portal, J.~M. Soler, The siesta
  method; developments and applicability, J. Phys. Cond. Mat. 20~(6) (2008)
  064208.

\bibitem{PBE}
J.~P. Perdew, K.~Burke, M.~Ernzerhof, Generalized gradient
approximation made
  simple, Phys. Rev. Lett. 77 (1996) 3865--3868.

\bibitem{Troullier91}
N.~Troullier, J.~L. Martins, Efficient pseudopotentials for
plane-wave
  calculations, Phys. Rev. B 43 (1991) 1993.

\bibitem{Monkhorst-Pack76}
H.~J. Monkhorst, J.~D. Pack, Special points for brillouin-zone
integrations,
  Phys. Rev. B 13 (1976) 5188--5192.

\bibitem{CGmethod}
M.~R. Hestenes, E.~Stiefel, Methods of conjugate gradients for
solving linear
  systems, J. Res. Natl. Bur. Stand. 49 (1952) 409--436.

\end{thebibliography}

\end{document}



\begin{frontmatter}



\title{{\normalsize Supplementary Material for:}\\
       Catalytic Formation of Narrow Nb Nanowires Inside
       Carbon Nanotubes}



\author{Dan~Liu} %

\author{David Tom\'{a}nek%
\corref{cor}} %
\ead{tomanek@pa.msu.edu}

\address{Physics and Astronomy Department,
            Michigan State University,
            East Lansing, Michigan 48824, USA}
\cortext[cor]{Corresponding author} %


\end{frontmatter}



\renewcommand\thesubsection{\Alph{subsection}}
\setcounter{subsection}{0} %
\renewcommand{\theequation}{S\arabic{equation}}
\setcounter{equation}{0} %
\renewcommand{\thefigure}{S\arabic{figure}}
\setcounter{figure}{0} %

\section{Reaction pathway}

As mentioned in the main manuscript, Nb is a refractory metal with
a high melting temperature $T_M=2750$~K that is close to the
decomposition temperature of carbon nanotubes (CNTs). Thus,
capillary filling, which had been used for other elements with a
much lower melting point, including Pb, can not be used for Nb.

Formation of Pb nanowires by capillary filling is a relatively
straight-forward process, as it only involves the condensation of
Pb atoms within the void inside a nanotube,
\begin{equation}
{\rm{Pb(g)}} \longrightarrow {\rm{Pb(s)}}\;.%
\label{EqS1}
\end{equation}
The reaction coordinate is the position of a single Pb atom in the
3-dimensional space and the reaction is thus easy to characterize.

For Nb, we propose a fundamentally different approach that yields
metallic Nb as one of the products of a complex reaction that
starts from the NbCl$_{3}$ molecular solid. As discussed in the
main manuscript, this reaction is
\begin{equation}
10{\rm{NbCl}}_3{\rm{(s)}} \longrightarrow %
3{\rm{Nb}}_{2}{\rm{Cl}}_{10}{\rm{(g)}} + 4{\rm{Nb(s)}} \;.%
\label{EqS2}
\end{equation}
With 40 atoms with 120 degrees of freedom involved, identifying
the optimum reaction path in the 120-dimensional configuration
space is practically impossible. Understanding this reaction is a
serious challenge in Inorganic Chemistry.

\begin{figure}[h]
\includegraphics[width=1.0\columnwidth]{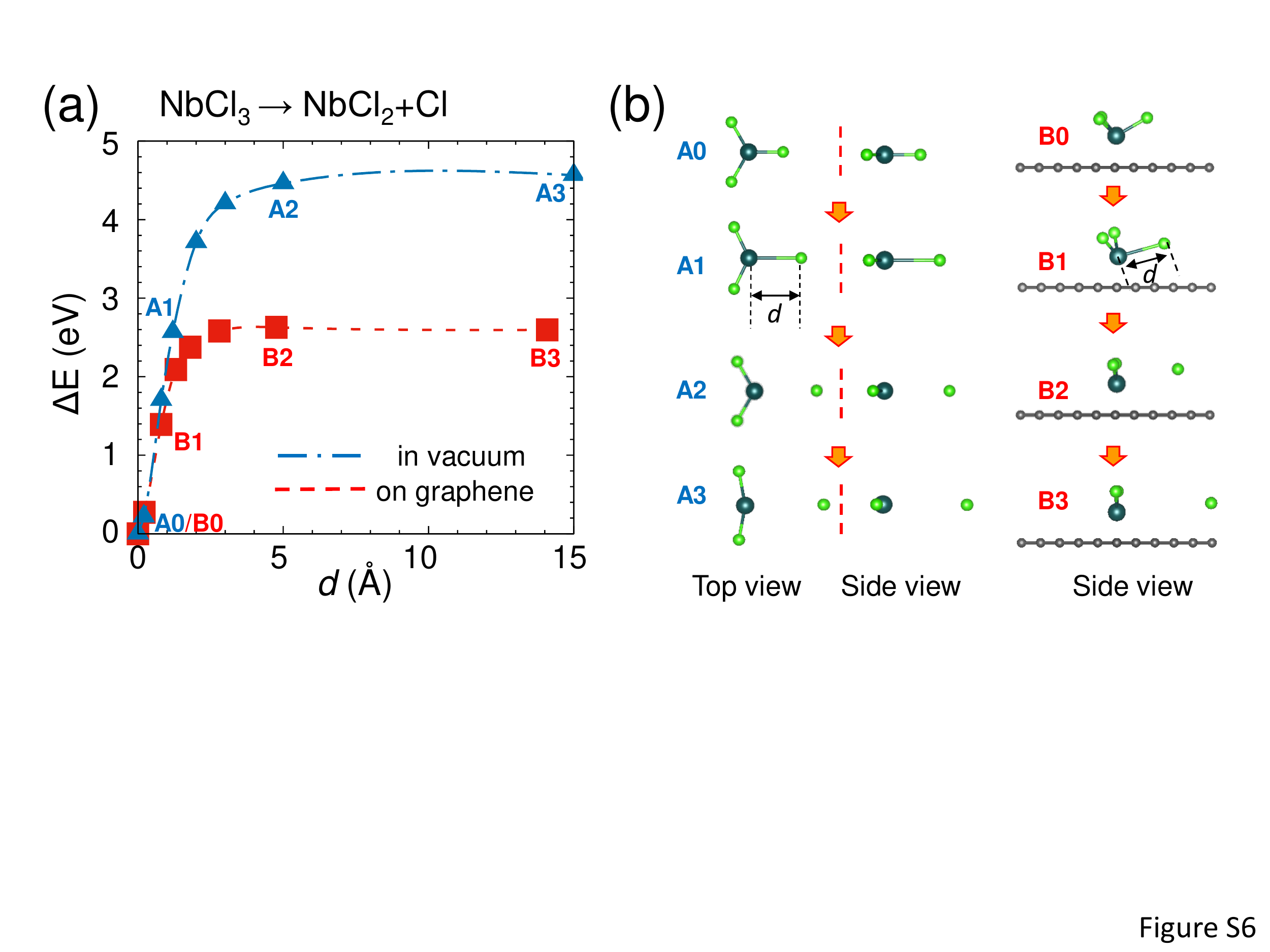}
\caption{Dissociation of NbCl$_3$ to NbCl$_2$ and Cl according to
Eq.~(\ref{EqS3}) in vacuum and on graphitic carbon. %
(a) Energetics of the reaction in vacuum (data points A0--A3,
connected by the blue dash-dotted line) and on graphitic carbon
(data points B0--B3, connected by the red dashed line). %
(b) Snap shots of the intermediate geometries. The reaction
coordinate is represented by $d$, the length of one particular
Nb-Cl bond. %
\label{figS1}}
\end{figure}
\begin{figure}[h!]
\includegraphics[width=1.0\columnwidth]{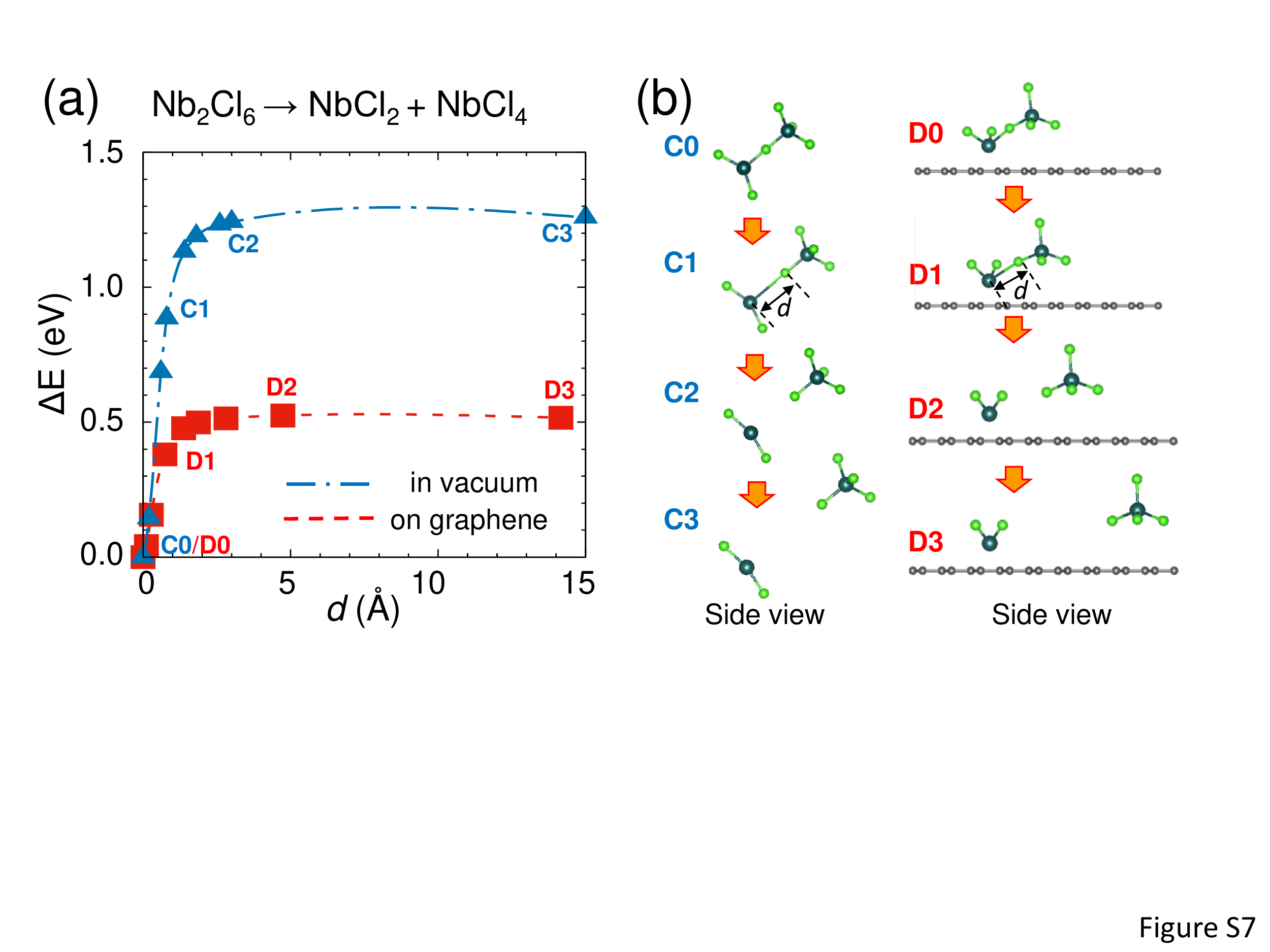}
\caption{Transformation of Nb$_2$Cl$_6$ to NbCl$_2$ and NbCl$_4$
according to Eq.~(\ref{EqS4}) in vacuum and on graphitic carbon. %
(a) Energetics of the reaction in vacuum (data points C0--C3,
connected by the blue dash-dotted line) and on graphitic carbon
(data points D0--D3, connected by the red dashed line). %
(b) Snap shots of the intermediate geometries. The reaction
coordinate is represented by $d$, the length of one particular
Nb-Cl bond. %
\label{figS2}}
\end{figure}

We have attempted to run molecular dynamics simulations reflecting
the reaction Eq.~(\ref{EqS2}) under high-pressure and
high-temperature conditions to get guidance regarding the likely
reaction path, but failed due to the limited simulation time. We
have concluded that locating intermediate structures and
transition states for a reaction as complex as in Eq.~(\ref{EqS2})
is not feasible computationally.

\begin{figure}[t!]
\includegraphics[width=1.0\columnwidth]{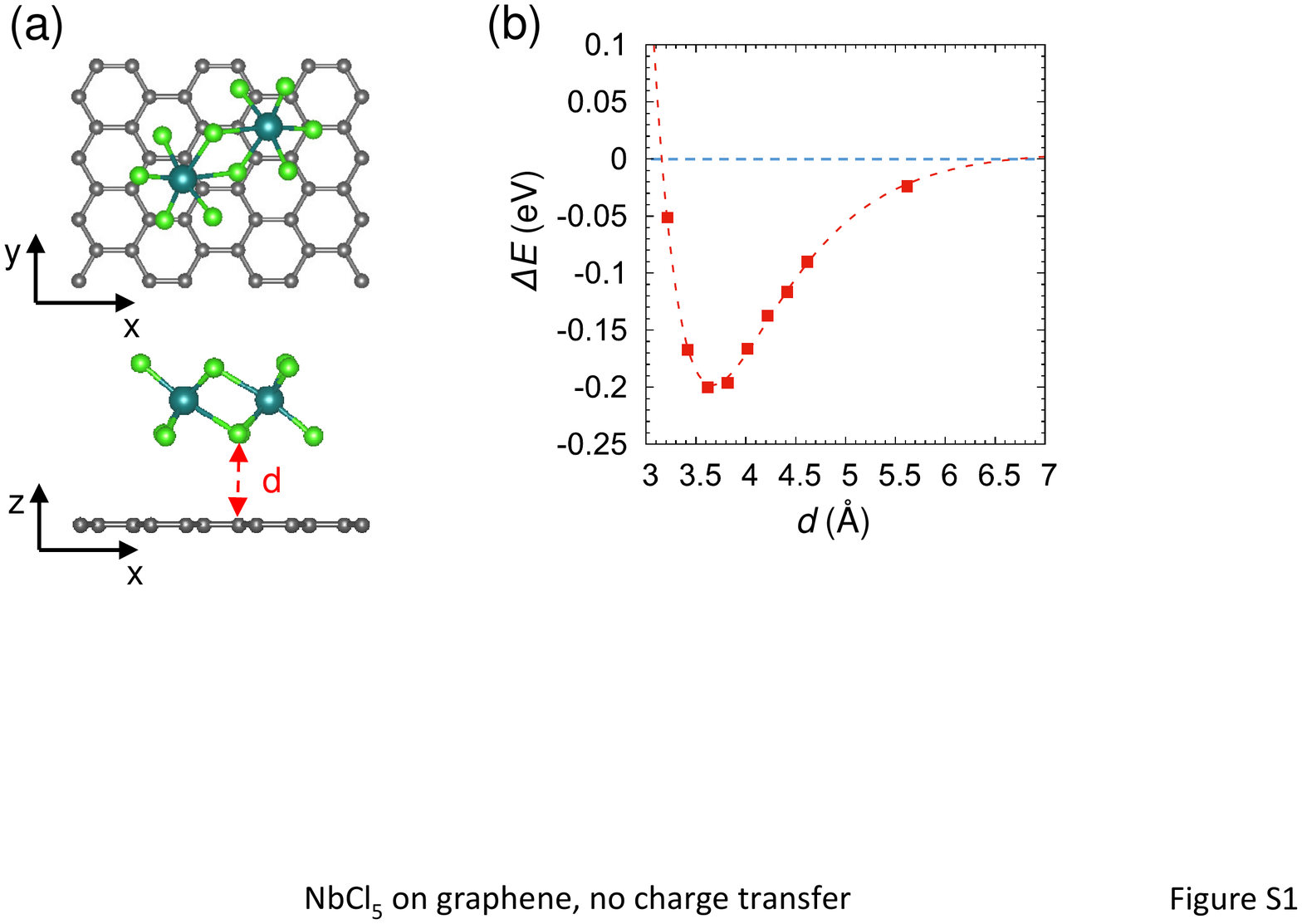}
\caption{Interaction of an Nb$_2$Cl$_{10}$ molecule
with a graphene monolayer.%
(a) Side and top views of the molecule at the distance $d$
    from graphene. %
(b) Interaction energy ${\Delta}E(d)$ between an Nb$_2$Cl$_{10}$
    molecule and graphene. %
\label{figS3}}
\end{figure}
\begin{figure}[h!]
\includegraphics[width=1.0\columnwidth]{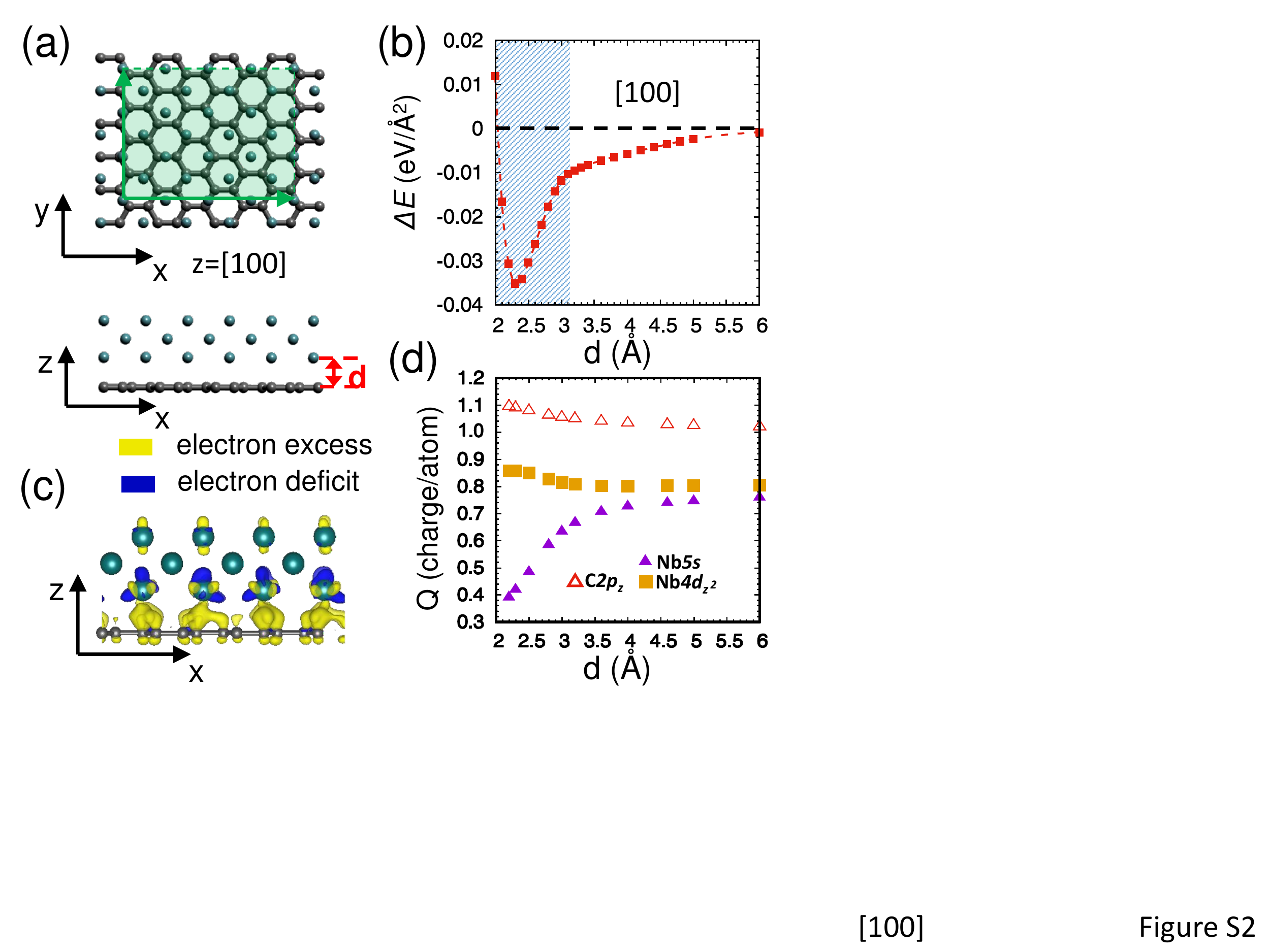}
\caption{Interaction of the Nb(100) surface with a graphene
         monolayer. %
(a) Top and side views of the atomic arrangement, with the
    supercell highlighted by red. %
(b) Interaction energy ${\Delta}E$ between the Nb(100) surface and
    graphene as a function of the separation distance $d$. %
(c) Charge redistribution caused by the interaction of Nb(100)
    with graphene, superposed with the atomic structure. %
    ${\Delta}\rho=\rho$(Nb/graphene)$-\rho$(Nb)$-\rho$(graphene)
    is shown by isosurfaces bounding regions of electron excess at
    $+4{\times}10^{-3}$~e/{\AA}$^3$ (yellow) and electron
    deficiency at $-4{\times}10^{-3}$~e/{\AA}$^3$ (blue). %
(d) Mulliken charge of specific atomic orbitals as a function of
    $d$.
\label{figS4}}
\end{figure}

Still, we can provide results of limited calculations for
processes that likely take place during the complex reaction in
Eq.~(\ref{EqS2}), both in vacuum and inside a carbon nanotube.

One reaction step that is likely to occur in this complex reaction
is the detachment of a chlorine atom from an NbCl$_{3}$ molecule
of the starting substance according to
\begin{equation}
{\rm{NbCl}}_3{\rm{(g)}} \longrightarrow %
{\rm{NbCl}}_2{\rm{(g)}} + {\rm{Cl(g)}} \;.%
\label{EqS3}
\end{equation}
The energetics of the corresponding endothermic dissociation
process, which may occur in vacuum or on graphitic carbon, is
presented in Fig.~\ref{figS1} along with the corresponding
geometries. Clearly, the dissociation energy is lowered
significantly in presence of graphitic carbon.

\begin{figure}[t]
\begin{center}
\includegraphics[width=1.0\columnwidth]{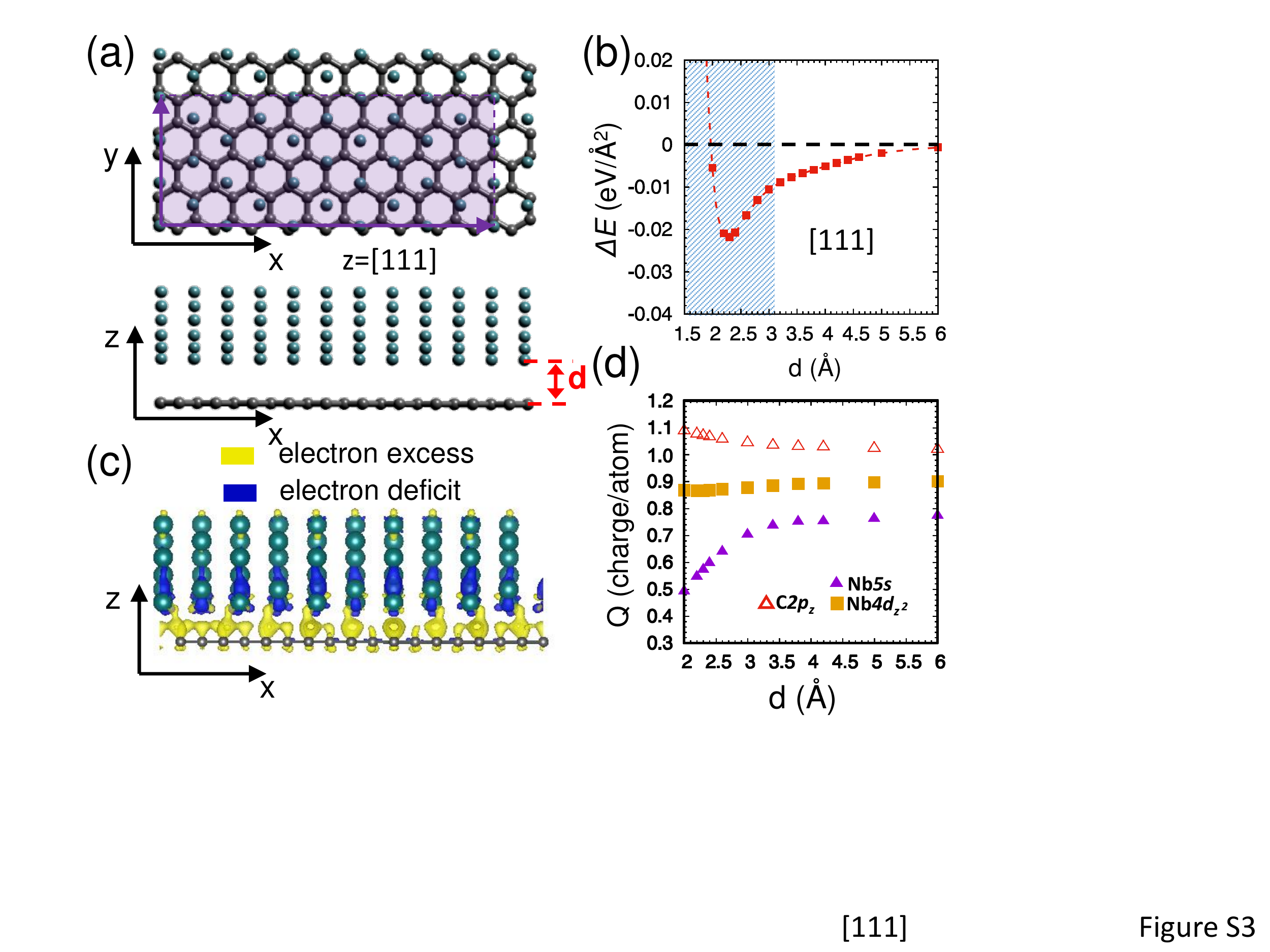}
\end{center}
\caption{Interaction of the Nb(111) surface with a graphene
         monolayer. %
(a) Top and side views of the atomic arrangement, with the
    supercell highlighted by red. %
(b) Interaction energy ${\Delta}E$ between the Nb(111) surface and
    graphene as a function of the separation distance $d$. %
(c) Charge redistribution caused by the interaction of Nb(111)
    with graphene, superposed with the atomic structure. %
    ${\Delta}\rho=\rho$(Nb/graphene)$-\rho$(Nb)$-\rho$(graphene)
    is shown by isosurfaces bounding regions of electron excess at
    $+4{\times}10^{-3}$~e/{\AA}$^3$ (yellow) and electron
    deficiency at $-4{\times}10^{-3}$~e/{\AA}$^3$ (blue). %
(d) Mulliken charge of specific atomic orbitals as a function of
    $d$.
\label{figS5}}
\end{figure}
\begin{figure}[t]
\includegraphics[width=1.0\columnwidth]{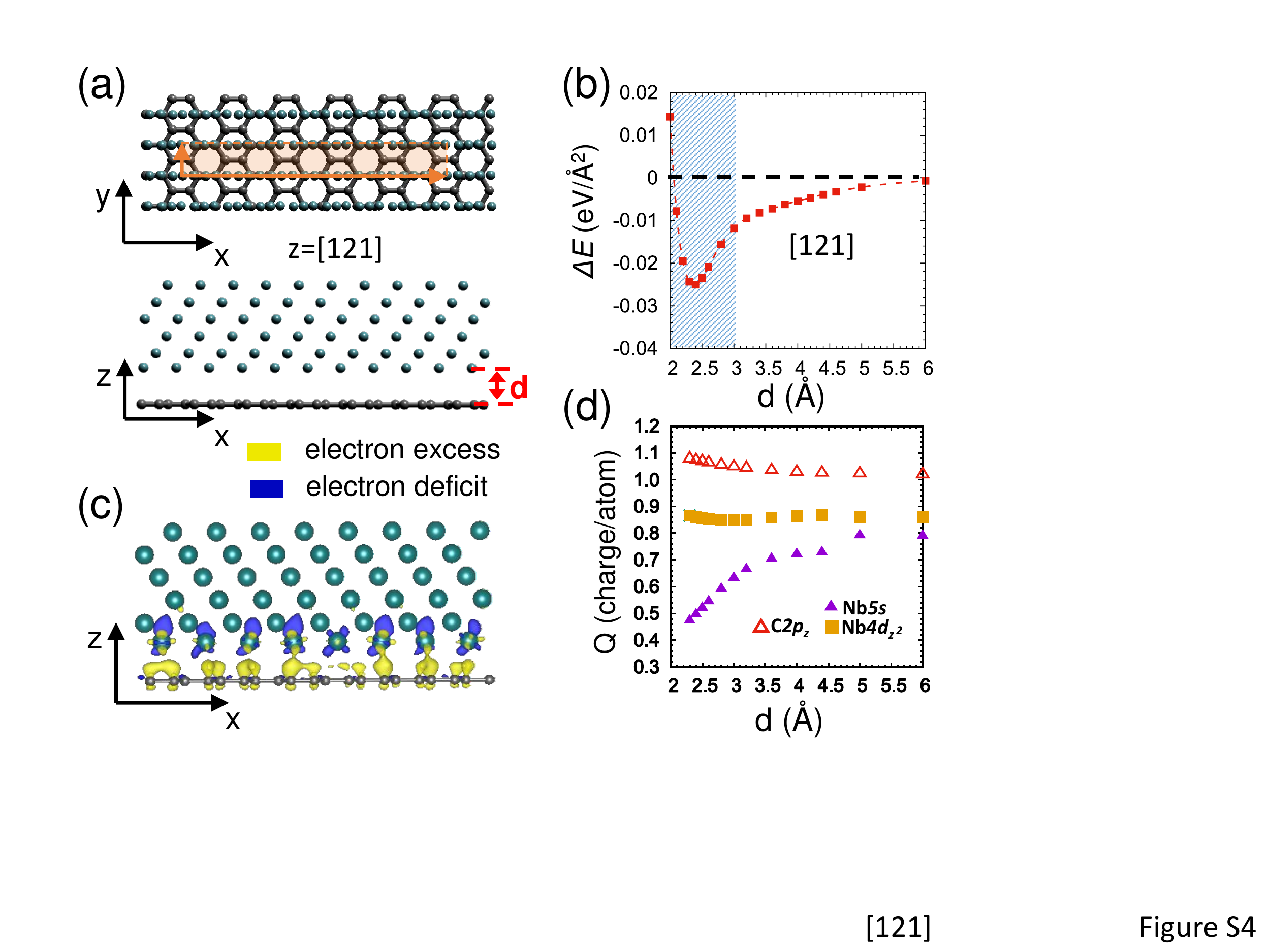}
\caption{Interaction of the Nb(121) surface with a graphene
         monolayer. %
(a) Top and side views of the atomic arrangement, with the
    supercell highlighted by red. %
(b) Interaction energy ${\Delta}E$ between the Nb(121) surface and
    graphene as a function of the separation distance $d$. %
(c) Charge redistribution caused by the interaction of Nb(121)
    with graphene, superposed with the atomic structure. %
    ${\Delta}\rho=\rho$(Nb/graphene)$-\rho$(Nb)$-\rho$(graphene)
    is shown by isosurfaces bounding regions of electron excess at
    $+4{\times}10^{-3}$~e/{\AA}$^3$ (yellow) and electron
    deficiency at $-4{\times}10^{-3}$~e/{\AA}$^3$ (blue). %
(d) Mulliken charge of specific atomic orbitals as a function of
    $d$.
\label{figS6}}
\end{figure}

Another reaction likely to occur during the conversion of
NbCl$_{3}$ to metallic Nb and other molecules is the dimerization
of NbCl$_{3}$ to Nb$_2$Cl$_{6}$, followed by simultaneous
detachment of a Cl atom from NbCl$_{3}$ while a new Nb-Cl bond is
being formed with another NbCl$_{3}$ molecule. One reaction that
reflects simultaneous dissociation and formation of an Nb-Cl bond
is
\begin{equation}
{\rm{Nb}}_{2}{\rm{Cl}}_{6}{\rm{(g)}} \longrightarrow %
{\rm{NbCl}}_2{\rm{(g)}} + {\rm{NbCl}}_4{\rm{(g)}} \;.%
\label{EqS4}
\end{equation}
The energetics of this endothermic dissociation process, which may
occur in vacuum or on graphitic carbon, is illustrated in
Fig.~\ref{figS2} along with the corresponding geometries..
Clearly, the energy barrier is lowered in presence of graphitic
carbon also for this reaction.

\section{%
Interaction of NbCl$_5$ with graphene}

NbCl$_5$ dimerizes to Nb$_2$Cl$_{10}$ molecules shown in
Figure~\ref{figS3}(a). In contrast to NbCl$_3$ discussed in the
main manuscript, there is only one shallow `physisorption' state
of Nb$_2$Cl$_{10}$ on graphene, as seen in Figure~\ref{figS3}(b).
The physisorption energy ${\Delta}E=-0.2$~eV is very small and the
charge transfer between the molecule and graphene is negligible.

\section{Interaction of Nb(100), Nb(111) and Nb(121) surfaces
               with graphene}

An analogy to the results for interaction of Nb(110) with
graphene, presented in Figure~3 of the main manuscript, we present
results for graphene interacting with Nb(100) in
Figure~\ref{figS4}, Nb(111) in Figure~\ref{figS5}, and Nb(121) in
Figure~\ref{figS6}.

\begin{figure}[h]
\includegraphics[width=1.0\columnwidth]{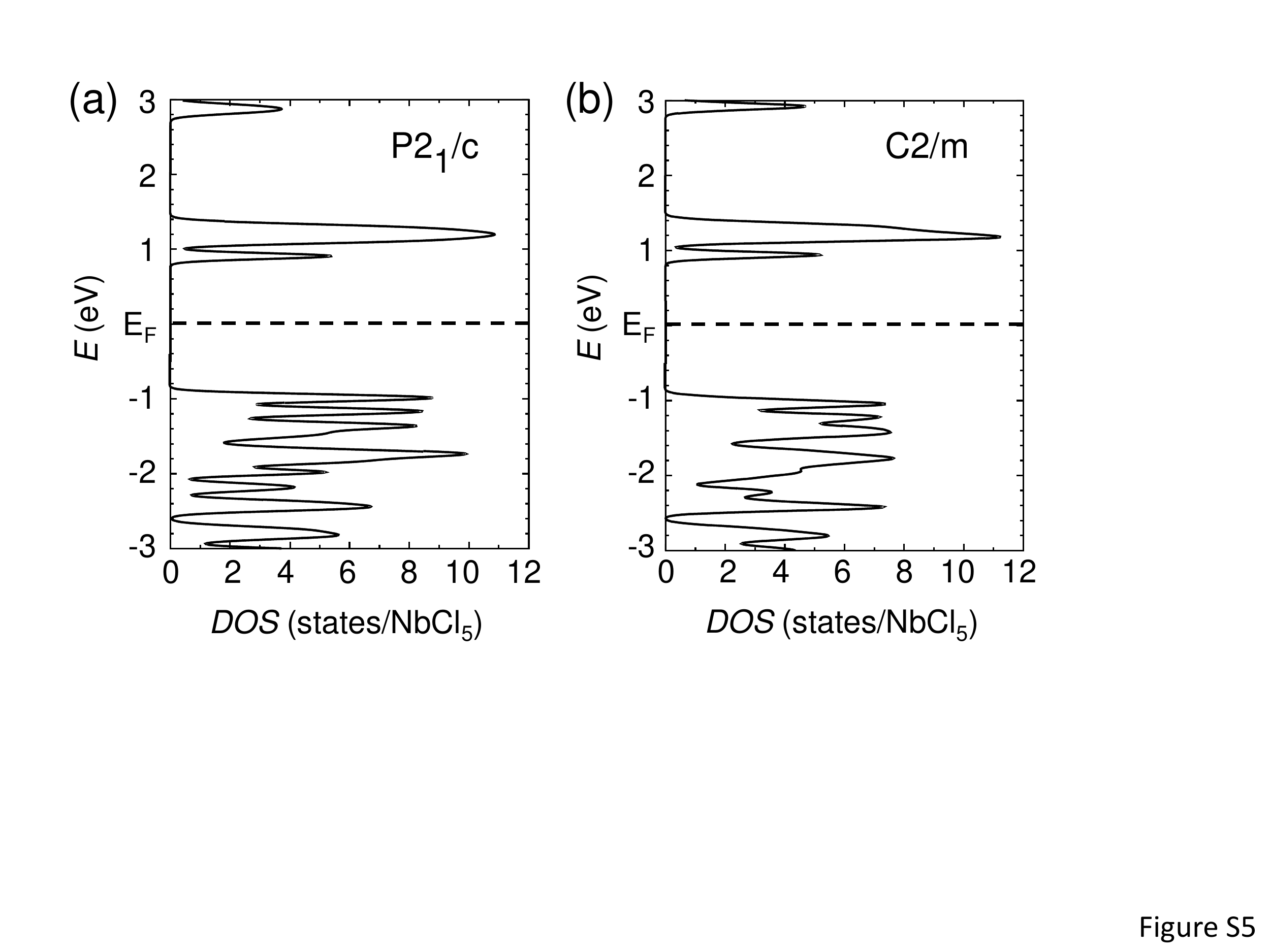}
\caption{Electronic density of states (DOS) of two monoclinic
         bulk allotropes of niobium pentachloride. %
(a) NbCl$_5$ with $P2_1/c$ symmetry and $24$ atoms in total per unit cell. %
(b) NbCl$_5$ with $C2/m$ symmetry and $72$ atoms in total per unit cell. %
Both crystals are wide-gap semiconductors. %
\vspace*{14cm} %
\label{figS7}}
\end{figure}

\section{Electronic structure of bulk NbCl$_5$}

We have optimized the bulk structure of two bulk allotropes of
NbCl$_5$ using density functional theory (DFT) as specified in the
main manuscript. We have found their energies to differ by
${lesssim}1$~meV per formula unit, indicating that the two
allotropes are equally stable. We present the DFT-based electronic
density of states (DOS) in Fig.~\ref{figS7}. In view of the fact
the DFT eigenvalues always underestimate the fundamental band gap,
both allotropes are clearly wide-gap semiconductors.